\documentclass[reprint,superscriptaddress,amsmath,amssymb,aps,prl]{revtex4-1}
\usepackage{graphicx}
\usepackage{placeins}
\usepackage{bm}
\usepackage{xcolor}

\begin{document}

\title{Controllable high-speed polariton waves in a PT-symmetric lattice}

\author{Xuekai Ma}
\email{xuekai.ma@gmail.com}
 \affiliation{Department of Physics and Center for Optoelectronics and Photonics Paderborn (CeOPP), Universit\"{a}t Paderborn, Warburger Strasse 100, 33098 Paderborn, Germany}
\author{Yaroslav Y. Kartashov}%
\affiliation{ 
Institute of Spectroscopy, Russian Academy of Sciences, Troitsk, Moscow, 108840, Russia
}%

\author{Tingge Gao}
\affiliation{%
Tianjin Key Laboratory of Molecular Optoelectronic Science, Institute of Molecular Plus, Tianjin University, Tianjin 300072, China
}%
\affiliation{%
Department of Physics, School of Science, Tianjin University, Tianjin 300072, China
}%

\author{Stefan Schumacher}
 \affiliation{Department of Physics and Center for Optoelectronics and Photonics Paderborn (CeOPP), Universit\"{a}t Paderborn, Warburger Strasse 100, 33098 Paderborn, Germany}
 \affiliation{College of Optical Sciences, University of Arizona, Tucson, AZ 85721, USA}
\date{\today}

\begin{abstract}
Parity-time (PT) symmetry gives rise to unusual phenomena in many physical systems, presently attracting a lot of attention. One essential and non-trivial task is the fabrication and design of the PT-symmetric lattices in different systems. Here we introduce a method to realize such a lattice in an exciton-polariton condensate in a planar semiconductor microcavity. We theoretically demonstrate that in the regime, where lattice profile is nearly PT-symmetric, a polariton wave can propagate at very high velocity resulting from the beating of a ground state condensate created in the lowest energy band at very small momentum and a condensate simultaneously created in higher energy states with large momentum. The spontaneous excitation of these two states in the nonlinear regime due to competition between multiple eigenmodes becomes possible since the spectrum of nearly PT-symmetric structure reveals practically identical amplification for Bloch waves from the entire Brillouin zone. There exists a wide velocity range for the resulting polariton wave. This velocity can be controlled by an additional coherent pulse carrying a specific momentum. We also discuss the breakup of the PT-symmetry when the polariton lifetime exceeds a certain threshold value.
\end{abstract}

\maketitle

\section{Introduction}
Quantum mechanics has shown tremendous success in describing microscale particles and even macroscopic objects. One of the physical axioms in quantum mechanics is that the Hamiltonian should be Hermitian so that the system has real eigenenergies and unitary time evolution. However, there exist many systems effectively described by non-Hermitian Hamiltonians~\cite{moiseyev2011non}, for example, quantum particles tunneling through a semi-transparent barrier or waves propagating in complex refractive index landscapes in optics. Generally the eigenenergies are complex valued for a non-Hermitian Hamiltonian. Recently, the physics of PT symmetry has attracted a lot of attention in various systems. Originally proposed by Bender and Boettcher ~\cite{bender1998real} (see reviews \cite{suchkov2016nonlinear,konotop2016nonlinear}), PT-symmetric systems are described by Hamiltonians that are invariant under the action of the PT operator, where P is the parity operator, and T is the time-reversal operator. In these systems the energy spectrum may be real despite the non-Hermiticity of the Hamiltonian describing the systems. Such a PT-symmetric system can experience a phase transition from an entirely real-valued eigenvalue spectrum to a complex-valued one when the ratio between the imaginary part and real part of the complex potential exceeds a certain critical value. This specific point, called symmetry-breaking point or exceptional point~\cite{heiss2001chirality}, occurs when both the eigenenergies and eigenvectors of the system coalesce.

Due to mathematical similarity between the Schr\"{o}dinger equation and paraxial wave equation describing propagation of light~\cite{el2007theory,abdullaev2011solitons}, photonic or optical systems can be used to study PT symmetry, where the non-Hermitian potential for light can be judiciously engineered by the fabrication of materials with complex refractive index. To achieve PT symmetry, the real part of the refractive index distribution has to be an even function in space and the imaginary part has to be an odd function. Then, in the PT-symmetric phase, the modes have symmetric modulus distributions and do not experience net gain or loss. Above the PT symmetry breaking point, one state experiences net gain and another net loss. At the phase transition point many counter-intuitive features are observed which cannot be found in Hermitian systems, such as unidirectional transport of light along a loss/gain modulated waveguide~\cite{feng2013experimental,feng2014single}, loss-induced suppression and revival of lasing~\cite{peng2014loss}, enhanced sensitivity of a microcavity sensor~\cite{hodaei2017enhanced,chen2017exceptional}, single mode lasing and the appearance of a vortex laser in a ring-shaped micro device~\cite{feng2014single,miao2016orbital} (more examples are found in the review~\cite{el2018non}).

\begin{figure*} 
\centering
\includegraphics[width=1.8\columnwidth]{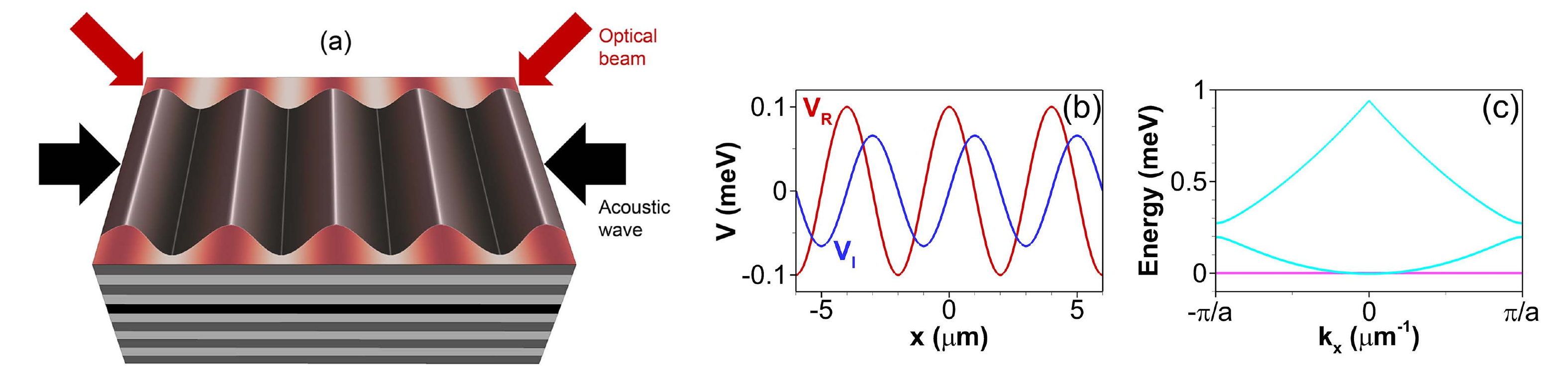}
\caption{(a) Sketch of a planar semiconductor quantum-well microcavity with excitation by a periodically modulated optical beam in the presence of a periodic external potential, e.g., created by two counter-propagating acoustic waves. (b) Real part ($V_\text{R}$) and imaginary part ($V_\text{I}$) of a complex external potential with $V_0=0.1$ meV. (c) Band structure for the complex periodic potential in (b) {with effective mass $m_{\text{eff}}=10^{-4}m_{\text{e}}$ ($m_{\text{e}}$ is the free electron mass)}. Real (cyan) and imaginary (pink) parts of the complex eigenvalues are shown.}\label{sketch}
\end{figure*}

Considerable progress has been made in the theoretical and experimental investigation of PT symmetric photonic systems. Our aim in this work is to show that PT-symmetric lattices can be realized in polariton systems, where they can lead to unusual dynamics of polariton condensates. We use exciton polaritons in a planar microcavity to study PT symmetry in a 1D periodic lattice under optical pumping. Exciton polaritons form due to the strong coupling between quantum well excitons and cavity photons, and they can undergo condensation~\cite{deng2002condensation,kasprzak2006bose} at much higher temperature~\cite{christopoulos2007room} than cold atoms. However, due to the spontaneous decay of polaritons, persistent optical pumping has to be used to sustain the population in polariton condensates. In this case, the system is intrinsically driven-dissipative and non-Hermitian~\cite{gao2015observation,gao2018chiral}. Polaritons have both photonic and excitonic components, that opens a wealth of opportunities to tailor the complex potential and investigate the nonlinear dynamics in polariton lattices~\cite{tanese2013polariton,cerda2013exciton,ostrovskaya2013self,ma2017creation}. Persistent Rabi oscillation~\cite{chestnov2016permanent} and multistability of polaritons have been already predicted in PT symmetric coupled microcavities~\cite{lien2015multistability}. However, the spatial evolution of polariton condensates in extended PT-symmetric lattices has never been proposed or explored, although it is apparent that such lattices may open new prospects for the manipulation of the polariton waves.

In the present paper we realize a PT-symmetric lattice for polariton condensates, using a spatially periodically modulated optical pump profile together with a periodically modulated external potential (for example induced by the interference of two counter-propagating acoustic waves), cf. sketch in Fig.~1. Here we demonstrate the general feasibility of an exactly PT-symmetric lattice. We also show that in the regime when the lattice profile only slightly deviates from the ideal PT-symmetric landscape, when the constant background in the periodic pump beam only slightly exceeds the threshold for condensation, the lattice provides practically identical gain for Bloch modes in the entire first Brillouin zone. The nonlinear competition between such growing modes excited with noisy input results in the formation of ground and excited polariton states at different momentum values. The resulting beating of these condensate fractions generates a high-speed polariton wave. The velocity of the resulting wave can be optically controlled in a wide range using an additional coherent pulse. We also demonstrate that the PT-symmetry can be broken as the lifetime of polariton increases, giving rise to an asymmetric spectrum.

\begin{figure*} [htp]
\centering
\includegraphics[width=1.8\columnwidth]{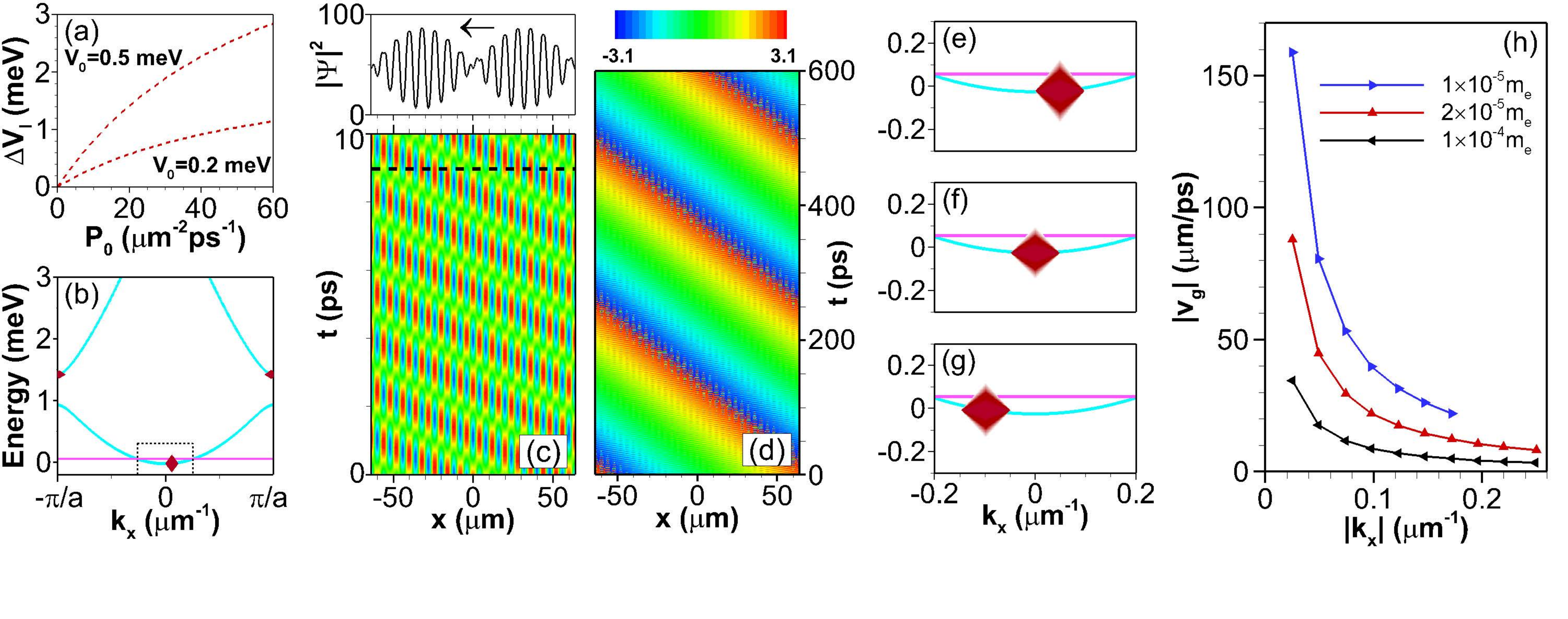}
\caption{(a) Deviation of the imaginary part of the potential per period from the corresponding exactly anti-symmetric distribution versus $P_0$ for different strengths of the external potential. (b) Real (cyan color) and imaginary (pink color) parts of the linear band structure for $V_0=0.5$ meV and $P_0=5~\mu$m$^{-2}$ps$^{-1}$ ($P_{thr}=60~\mu$m$^{-2}$ps$^{-1}$). Included are the condensate spectra calculated from the dynamics in (c) and (d). Time evolution of (c) the density of the polariton condensate and (d) the corresponding phase on long time scales. The upper panel in (c) shows a 1D slice through the density ($\mu$m$^{-2}$) at $t=9$ ps as indicated by the dashed line. (e)-(g) The lowest energy band including examples of the ground state spectra of the final polariton distribution emerging in three cases of randomly different noisy input. The spectra in the higher bands are the same as in (b). (e) Enlarged view of the dashed box region in (b). (h) Velocities of polariton waves depending on ground state polariton momentum for different effective masses and fixed external potential with $V_0=0.5$ meV. For (a)-(g) $m_{\text{eff}}=2\times10^{-5}m_\text{e}$.}\label{moving}
\end{figure*}

\section{Theoretical model}
The dynamics of a polariton condensate in semiconductor microcavities at the bottom of the lower-polariton branch can be described by a driven-dissipative Gross-Pitaevskii (GP) model, coupled to an equation for the density of the exciton reservoir~\cite{wouters2007excitations}:
\begin{equation}\label{e1}
\begin{aligned}
i\hbar\frac{\partial\Psi(\mathbf{r},t)}{\partial t}&=\left[-\frac{\hbar^2}{2m_{\text{eff}}}\nabla_\bot^2-i\hbar\frac{\gamma_\text{c}}{2}+g_\text{c}|\Psi(\mathbf{r},t)|^2 \right.\\
&+\left.\left(g_\text{r}+i\hbar\frac{R}{2}\right)n(\mathbf{r},t)+V_\text{e}(\mathbf{r},t)\right]\Psi(\mathbf{r},t)
\end{aligned}
\end{equation}
\begin{equation}\label{e2}
\frac{\partial n(\mathbf{r},t)}{\partial t}=\left[-\gamma_r-R|\Psi(\mathbf{r},t)|^2\right]n(\mathbf{r},t)+P(\mathbf{r},t)\,.
\end{equation}
Here $\Psi(\mathbf{r},t)$ is the coherent polariton field and $n(\mathbf{r},t)$ is the exciton reservoir density. The effective mass of polaritons is $m_{\text{eff}}$. Due to the finite lifetime of polaritons, the condensate has a decay rate $\gamma_\text{c}=0.2$ ps$^{-1}$, and the reservoir decays with $\gamma_\text{r}=0.3$ ps$^{-1}$. The polariton condensate is replenished by the coupling to the reservoir density $n(\mathbf{r},t)$ with rate $R=0.001$ $\mu$m$^{2}$ ps$^{-1}$, while the reservoir is excited by an incoherent pump $P(\mathbf{r},t)$. The interaction strength between polaritons is given by $g_\text{c}=10^{-5}$ meV $\mu$m$^{2}$ and between polaritons and reservoir by $g_\text{r}=2g_\text{c}$. In different materials the interaction strengths and consequent nonlinearities can be significantly different, with typical interaction strenghts of 1-10 $\mu$eV $\mu$m$^2$ in inorganic materials~\cite{ferrier2011interactions,steger2015slow} and on the order of $10^{-3}$ $\mu$eV $\mu$m$^2$ in some organic materials~\cite{daskalakis2015spatial,sanvitto2016road,lerario2017room}. $V_\text{e}$ is an external potential which can be fabricated by different techniques~\cite{balili2007bose,lai2007coherent,wertz2010spontaneous,kim2013exciton}.

\section{PT-symmetric lattice}
Considering the 1D case, for our system the total external potential seen by the coherent condensate is given by
\begin{equation}\label{TotalPotential}
\begin{aligned}
V(x)&=V_{\text{R}}(x)+iV_{\text{I}}(x) \\
&=g_\text{r} n(x)+V_\text{e}(x)+i\hbar\left(\frac{R}{2}n(x)-\frac{\gamma_c}{2}\right),
\end{aligned}
\end{equation}
with the real potential $V_{\text{R}}(x)\equiv g_\text{r} n(x)+V_\text{e}(x)$ and the imaginary potential $V_{\text{I}}(x)\equiv\hbar\left(Rn(x)-\gamma_\text{c}\right)/2$. In general, both the real potential and the imaginary potential are related to the pump profile, resulting in that they always have the same symmetry in the absence of the external potential $V_\text{e}(x)$. However, when the external potential is present and much stronger than the reservoir-induced one, i.e., $V_\text{e}(x)\gg g_\text{r} n(x)$, the situation can change such that $V_{\text{R}}$ and $V_{\text{I}}$ have different symmetries.
Here, we use $V_\text{e}(x)=V_0\cos(2\pi x/a)$, which can for example be realized by interference of counter propagating acoustic waves [see Fig. \ref{sketch}(a)]~\cite{de2006phonon}. To obtain an antisymmetric profile of the imaginary part of the potential, we apply a spatially periodic pump satisfying
\begin{equation}\label{IncohPumpExpand2}
P(x)=\frac{A_0}{2}\sin(2\pi x/a)+\frac{A_0}{2}+P_0\,,
\end{equation}
with lattice constant $a=4$ $\mu$m. Such an intensity profile can be created using three continuous wave optical plane waves with different angles of incidence. We note that for spatially homogeneous excitation the pump threshold for condensation is $P_{\text{thr}}=\gamma_\text{c}\gamma_\text{r}/R$. Approaching the linear case, that is when the pump intensity is just above the condensation threshold, the steady-state reservoir density satisfies
\begin{equation}\label{ReservoirSimplify}
n(x)\simeq\frac{P(x)}{\gamma_\text{r}}.
\end{equation}
Substituting Eqs.~\eqref{IncohPumpExpand2} and \eqref{ReservoirSimplify} into Eq.~\eqref{TotalPotential}, the imaginary potential can then be written as
\begin{equation}\label{PotentialIM}
V_{\text{I}}(x)=\frac{\hbar}{2}\left[\frac{RA_0}{2\gamma_\text{r}}\sin(2\pi x/a)+\frac{R}{2\gamma_\text{r}}(A_0+2P_0)-\gamma_\text{c}\right],
\end{equation}
which is anti-symmetric when the constant loss rate, $\gamma_\text{c}$, is compensated by the spatially homogeneous part of the pump, i.e.,
\begin{equation}\label{ralationAnti}
\frac{A_0}{2}+P_0=\frac{\gamma_\text{c}\gamma_\text{r}}{R}=P_{\text{thr}}.
\end{equation}
In the following we set $A_0\equiv2P_{\text{thr}}$ and adjust the pump intensity by changing $P_0$.

\begin{figure*} 
\centering
\includegraphics[width=1.8\columnwidth]{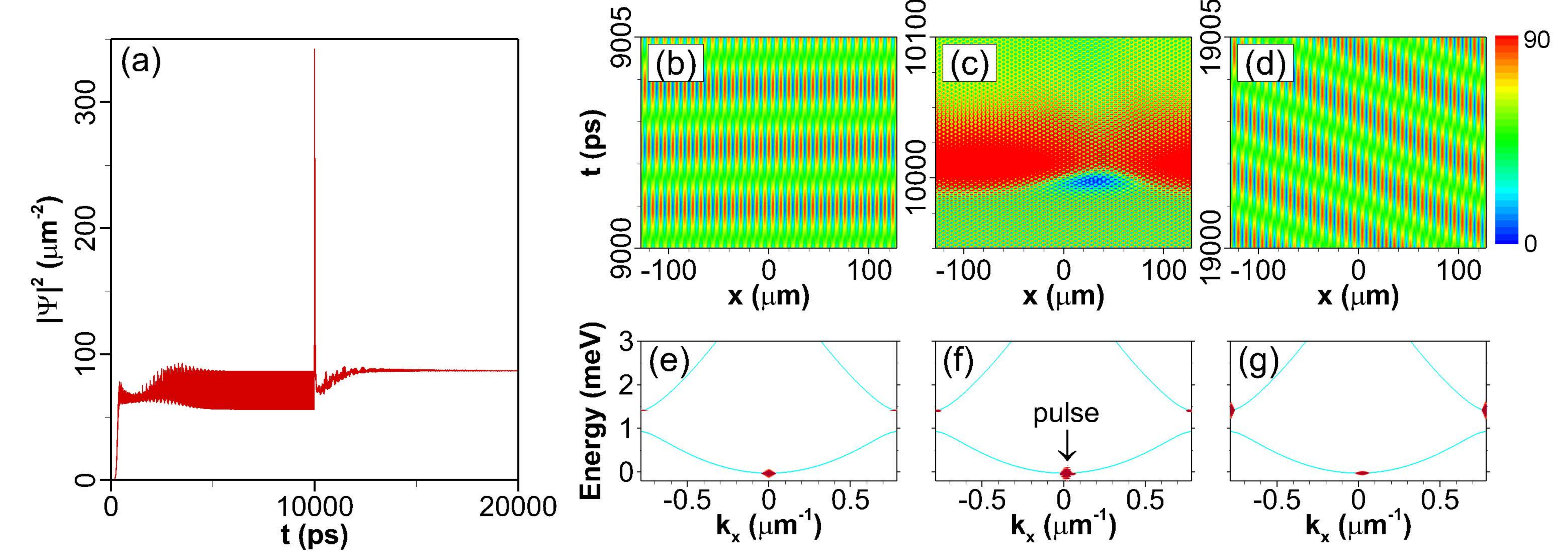}
\caption{Optical control of the velocity. (a) Time evolution of the peak density of the condensate with a coherent pulse applied at $t=10000$ ps. (b)-(d) Time evolution of spatial profiles at different points in time in (a). (e)-(g) Real part of the linear band structure and spectra for (b)-(d), respectively, with $V_0=0.5$  meV, $P_0=5~\mu$m$^{-2}$ps$^{-1}$, and $m_{\text{eff}}=2\times10^{-5}m_\text{e}$.}\label{switching}
\end{figure*}

We start with the simplest case of $P_0=0$, corresponding to exact PT-symmetric lattice configuration. The distributions of the resulting antisymmetric imaginary potential and symmetric real potential, which satisfy the PT-symmetry condition, are shown in Fig. \ref{sketch}(b). In this case the system possesses only real-valued eigenvalues (for the selected amplitude $A_0$) as shown in Fig.~\ref{sketch}(c).

To create larger condensate densities, such that the nonlinearity must be considered, an external pump with $P_0>0$ is required. Then, nonlinearity affects the distribution of the real potential as well as the imaginary potential through depletion of the reservoir. It can be seen from Eqs.~\eqref{e2} and \eqref{TotalPotential} that if the external potential ($V_\text{e}$) and the loss of the reservoir ($\gamma_\text{r}$) are not large enough, the total potential cannot satisfy a precise PT symmetry anymore. The imaginary potential will be further away from being anti-symmetric when the density of the condensate is larger. The resulting deviation of the imaginary potentials from a precise anti-symmetric distribution [$\tilde{V_I}=\tilde{V_{I0}}\sin(2\pi x/a)$ with $\tilde{V_{I0}}=\hbar RA_0/4\gamma_\text{r}$], defined as $\Delta V_I=\int|V_I-\tilde{V_I}|dx$, over one period is shown in Fig.~\ref{moving}(a). It is clear that the deviation increases with the intensity of the additional plane wave and the strength of the external potential. We note, however, that in the parameter range shown in Fig.~\ref{moving}(a) the deviation does not exceed 0.1\% of the magnitude of the imaginary part that means that the profile of such lattice remains close to the ideal PT-symmetric configuration. As a consequence in this regime the spectrum of the structure enables excitation of the high-speed polariton waves.


\begin{figure*}
\centering
\includegraphics[width=1.8\columnwidth]{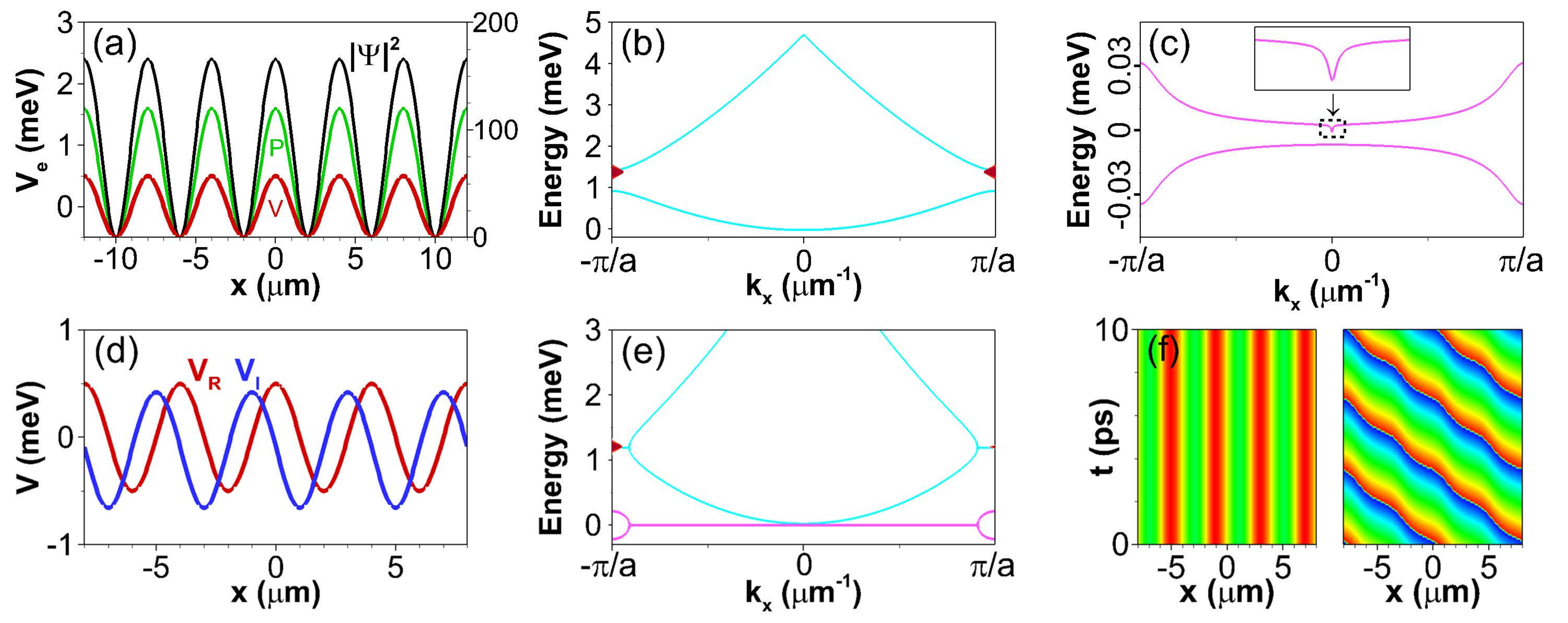}
\caption{Non-PT symmetry lattices and PT symmetry breaking. (a) Periodic distributions of external potential $V$ (scale on the left), pump profile $P$, and steady-state solution $|\psi|^2$ (scale on the right in $\mu$m$^{-2}$). (b) Real part of the linear band structure and the spectrum of the solution in (a). (c) Imaginary part of the linear band structure in (b). (d) Real and imaginary parts of the potential for a polariton lifetime $\gamma_\text{c}=2$ ps$^{-1}$, calculated from Eq.~\eqref{TotalPotential} with $n(x)=0$. (e) Real (cyan) and imaginary (pink) parts of the linear band structure for (d) including the polariton spectrum for (f). (f) Time evolution of the density and the phase of a stationary solution in the presence of the complex potential (d). Here, $V_0=0.5$ meV and $m_{\text{eff}}=2\times10^{-5}m_\text{e}$.}\label{nonPT}
\end{figure*}

As shown in Fig. \ref{moving}(b), when for $P_0>0$ the condensation occurs from noisy initial conditions and eventually reaches stationary behaviour, condensate fractions form at the bottoms of both bands. In Fig. \ref{moving}(b) the spectrum of the emerged condensate is superimposed on the eigenvalue spectrum of the nearly PT-symmetric linear lattice calculated at $P_0>0$ using the expression \eqref{TotalPotential} and assuming $|\Psi(x)|^2=0$. The condensate in the higher eigenstate is $k$-symmetric, while the condensate in the ground state carries a finite momentum, i.e., the spectrum slightly shifts from $k=0$ to finite $k$. We state that this is not an artifact, this also occurs in a very large calculation domain with an intensity filter to the periodic pump to avoid the influence of the boundaries. One can see from Fig. \ref{moving}(b) that the small deviation of the lattice profile from exact PT-symmetric landscape caused by $P_0$ results in the imaginary part of the eigenvalue being lifted from zero to a small positive value which is practically identical for all modes from the first Brillouin zone. Subsequent nonlinear competition between these modes excited by the initial noisy input typically results in the output containing excitations from two different bands. For an initial noise with zero average momentum, the fraction of the condensate belonging to the lower band typically has zero average momentum $k_x$ as shown in Fig. \ref{moving}(f). In contrast, when initial noisy input carries nonzero momentum, the output states can be generated that are concentrated, within the lower band, around nonzero momentum values, as shown in Figs. \ref{moving}(e) and \ref{moving}(g). It is worth mentioning that a homogeneous solution with a finite $k_x$, corresponding to the sole ground state, is unstable and then switches to the $k_x=0$ solution~\cite{ma2017creation}. Therefore, the presence of the excited state obvious in Fig. \ref{moving}(b) that competes with the ground state at finite $k_x$ value in nearly PT-symmetric structure seems to be essential for stabilization of the latter. Coexistence of the ground and the excited states with different momenta can also be attributed to the nonlinear character of gain, referring to the spontaneous scattering term $R|\Psi|^2$ in Eq. \eqref{e2}~\cite{bobrovska2017dynamical,estrecho2018single,ma2018vortex}. Interestingly, because of the interplay of the higher and ground state contributions, a high-speed polariton wave forms spontaneously with velocity of $\sim$44 $\mu$m/ps that moves to the left [see Fig. \ref{moving}(c)]. Here the velocity of this wave can be calculated by
\begin{equation}\label{velocity}
v_\text{g}=\Delta\omega/\Delta k_{x}=\Delta E/(\hbar\Delta k_x)\,,
\end{equation}
with the energy difference of the two contributing states $\Delta E\simeq-1.44$ meV and the momentum difference $\Delta k_x\simeq0.05$ $\mu$m$^{-1}$ from Fig.~\ref{moving}(c).

Figure~\ref{moving}(h) shows the dependence of the velocity of the resulting polariton wave on the momentum of the emerged ground state condensate. Increasing the momentum $k_x$ of the ground state condensate simultaneously causes a shift of its energy to higher value. When the effective mass of a polariton condensate is larger, the wave propagates slower than that with a smaller effective mass. This is because the energy difference between the ground and the excited states for a larger effective mass is smaller than for a smaller effective mass. Likewise, if the polaritons have a small effective mass ($10^{-5}m_{\text{e}}$), the velocity of the wave can be very high, more than 150 $\mu m$/ps at very small $k_x$ ($k_x=0.025$ $\mu$m$^{-1}$). From Fig.~\ref{moving}(h) we can also conclude that there is a very wide velocity range that can be obtained in a very small $k_x$ window.

Noisy initial conditions as discussed above can not easily be controlled. In the following we would like to explore the possibility to give the polariton condensate a desired momentum by application of a coherent pulse after the initial formation of the condensate. The results are shown in Fig.~\ref{switching}. From initial noise a symmetric oscillating state [Figs.~\ref{switching}(a) and \ref{switching}(b)] is created, where the ground state has a stronger contribution than the excited states [Fig. \ref{switching}(e)]. At $t=10000\,\mathrm{ps}$ we apply a short $10\,\mathrm{ps}$ coherent pulse, which is resonant with the lowest band, with a broad (near homogenous) spatial distribution. This pulse that carries a finite momentum, switches the condensate state from the symmetric initial oscillation to a unidirectionally propagating state. The velocity of the resulting polariton wave is related to the finite momentum carried by the coherent pulse. For example, a coherent pulse with $k_x=0.025$ $\mu$m$^{-1}$ [Figs. \ref{switching}(c) and \ref{switching}(f)] leads to the generation of an additional condensate at $k_x=0.025$ $\mu$m$^{-1}$ [Figs. \ref{switching}(d) and \ref{switching}(g)]. After the pulse is gone, the entire condensate in the ground state is permanently switched to the state with momentum dictated by the coherent pulse.

\section{non-PT-symmetric lattice}
To prove the coexistence of the stable ground state condensate around $k=0$ and the excited state at the boundaries of the Brillouin zone supported by PT-symmetry, we change the pump from spatially anti-symmetric to symmetric. The result for a typical symmetric pump is shown in Fig.~\ref{nonPT}(a), which only allows non-moving periodic solutions for the condensate. Figure~\ref{nonPT}(b) shows the corresponding band structure and condensate spectrum. The change of the pump obviously does not influence the real part of the band structure, while the imaginary part in Fig.~\ref{nonPT}(c) is strongly altered. In this case, during the initial formation of the condensate from initial noise, it simultaneously appears in the states at the center and at the boundaries of the Brillouin zone. However, the states at $k_x=0$ are quickly damped out due to the deep sink at $k=0$ [see the insert of Fig.~\ref{nonPT}(c)]. On long time scales only condensate in the states at the boundaries of the Brillouin remains as in Fig.~\ref{nonPT}(b). That is why in a non PT-symmetric case and for pump intensity close to the threshold two condensate fractions can only simultaneously be stable when they have the same magnitude of momentum~\cite{ma2015oscillatory}. A recent experimental work~\cite{baboux2018unstable} demonstrates that in the non-PT-symmetry scenario only one of the two states has been observed in one measurement when the pump intensity is just above the threshold, which agrees very well with our numerical finding. Notice that when the pump intensity is strong the phase transition of polariton condensates from $k_x=\pm\pi/a$ state to the ground state can also be observed~\cite{lai2007coherent,zhang2015weak}.


Previously, it has been mentioned that even in the PT-symmetric case the eigenenergies can become complex when the ratio of the amplitudes of the imaginary and real potentials exceeds a certain threshold value~\cite{heiss2001chirality}. In the present system, the depth of the imaginary potential in Eq.~\eqref{PotentialIM} can be adjusted by changing the rate $R/\gamma_\text{r}$ or $\gamma_\text{c}$. Here, we investigate PT-symmetry breaking by changing $\gamma_\text{c}$ with fixed $R/\gamma_\text{r}$. Increasing $\gamma_\text{c}$ leads to a stronger imaginary potential, significantly modifying also the real part of the band structure.
As $\gamma_\text{c}$ becomes larger at the point where the imaginary potential gets stronger than the real potential [Fig. \ref{nonPT}(d)], the real parts of the first and second band begin to merge at the boundaries of the first Brillouin zone [Fig. \ref{nonPT}(e)]. At the same time the imaginary parts of the band structure start to split, which shows the onset of PT-symmetry breaking. In this case, the condensate solution [Fig. \ref{nonPT}(f)] carries a homogeneous background and its density distribution does not change over time, while the phase velocity is non-zero. This coincides with a highly asymmetric spectral distribution in momentum space as shown in Fig. \ref{nonPT}(e).

\section{Conclusion}
In conclusion, we have introduced a method to realize a PT-symmetric lattice in polariton condensates. The method is based on a spatially periodic pump acting in a periodic external potential. When PT-symmetry is realized, a spatially propagating wave of polariton condensates is observed with a velocity related to the momentum of the condensate formed in the ground state. This velocity can be optically controlled to have a specific value inside a wide velocity range. At the onset of PT-symmetry breaking, where the system's eigenvalues become complex again, an asymmetric condensate solution is formed with stationary density, but non-zero phase velocity.

\section{acknowledgments}
This work was supported by the Deutsche Forschungsgemeinschaft (DFG) through the collaborative research center TRR142 (project A04, grant No. 231447078) and Heisenberg program (grant No. 270619725) and by the Paderborn Center for Parallel Computing, PC$^2$. X.M. further acknowledges support from the National Natural Science Foundation of China (grant No. 11804064).

\nocite{*}
%

\end{document}